\documentstyle{article}

 \normalmarginpar
\twocolumn
 \makeatletter
 \gdef\@proofbox{\relax}
 \long\def\proofbox#1{\gdef\@proofbox{#1}}
 \proofbox{\small{\sl PhysComp96}\\\ifx\UNDEF\@fullpaper
     Extended abstract\else Full paper\fi\\Draft, \@date}

 \gdef\fullpaper{\gdef\@fullpaper{}}

 \def\affil#1{\\{\small#1\par}}
 \gdef\@author{John Doe1\affil{No-Name University, Shipping Dept.}}
 \long\def\author#1{\gdef\@author{#1}}

 \gdef\@abstract{}
 \long\def\abstract#1{\gdef\@abstract{#1}}

\def\@maketitle{\newpage\leavevmode
  \begin{minipage}[t]{0.30\textwidth}
    \hrule height0pt
    \raggedright
    \mbox{}\par
    \@proofbox
  \end{minipage}\relax
  \begin{minipage}[t]{0.70\textwidth}
    \hrule height0pt
    \raggedleft
    \LARGE\@title\par
    \vskip4pt
    \large\@author
  \end{minipage}
  \vskip8pt
  \ifx\@abstract\@empty\else{\vskip.5em\leftskip1.5in\parskip4pt\small\@abstract\par\vskip.5em}\fi
  \rule{\textwidth}{0.4pt}
  \vskip16pt}

\DeclareRobustCommand\em
        {\@nomath\em \ifdim \fontdimen\@ne\font >\z@
                       \upshape \else \slshape \fi}

\def\@begintheorem#1#2{\sl \trivlist \item[\hskip \labelsep{\bf #1\ #2}]}
\def\@opargbegintheorem#1#2#3{\sl \trivlist
     \item[\hskip \labelsep{\bf #1\ #2\ (#3)}]}

 \newcommand{\sectlabel}[1]{\label{sect.#1}}

  \def\@arabic#1{\number #1} 

\long\def\@makecaption#1#2{
	\vskip\abovecaptionskip
	\sbox\@tempboxa{{\small #1: #2}}%
	\ifdim\wd\@tempboxa>\hsize
	    {\small #1: #2\par}
	\else
	   \global\@minipagefalse
	   \hbox to\hsize{\hfil\box\@tempboxa\hfil}
	\fi
	\vskip \belowcaptionskip}

\def\figstrut#1{\hbox to\linewidth{\vrule height#1\hfill}}

\def\comppad{\thinspace}
\def\comp{\comppad\begingroup \tt \let\do\@makeother \dospecials 
          \@ifstar{\@scomp}{\@comp}}
\def\@scomp#1{\def\@tempa ##1#1{##1\endgroup\comppad}\@tempa}
\def\@comp{\obeyspaces \frenchspacing \@scomp}

\makeatother

\title{Information Gain vs.\ State Disturbance \\ in Quantum Theory\bigskip}
\author{Christopher A. Fuchs\thanks{This work was supported in part by NSERC.}
\affil{D\'{e}partement IRO\\ Universit\'{e} de
Montr\'{e}al\\ C.~P.  6128, Succursale centre-ville\\ Montr\'eal,
Qu\'ebec, Canada H3C 3J7}}
\date{11 May 1996}

\abstract{The engine that powers quantum cryptography is the principle that
there are no physical means for gathering information about the identity of
a quantum system's state (when it is known to be prepared in one of a set
of nonorthogonal states) without disturbing the system in a statistically
detectable way.  This situation is often mistakenly described as a consequence 
of the ``Heisenberg uncertainty principle.''  A more accurate account is that it
is a unique feature of quantum phenomena that rests ultimately on the Hilbert
space structure of the theory {\it along\/} with the fact that time evolutions 
for isolated systems are unitary.  In this paper we shall explore several 
aspects of the {\it information/disturbance principle\/} in an attempt to make
it firmly quantitative and flesh out its significance for quantum theory as a 
whole.}

\begin{document}

\maketitle

\section{Introduction and Model}

Suppose an observer obtains a quantum system secretly prepared in one of two
nonorthogonal pure quantum states.  Quantum theory dictates that there is no 
measurement he can use to certify which of the two states was actually prepared.  This is well known \cite[and refs]{Fuchs96a}.  A simple, but less 
recognized, corollary is that no interaction used for performing such an
information-gathering measurement can leave both states unchanged in the
process \cite{Bennett92a}. If the observer could completely regenerate the 
unknown quantum state after measurement, then---by making further nondisturbing 
information-gathering measurements on it---he would be able eventually to infer 
the state's identity after all.

This consistency argument is enough to establish a tension between information
gain and disturbance in quantum theory.  What it does not capture, however, is
the extent of the tradeoff between these two quantities.  In this talk, we shall lay the groundwork for a quantitative study that goes beyond the qualitative 
nature of this tension.  Namely, we will show how to capture in a formal way
the idea that, depending upon the particular measurement interaction, there can 
be a tradeoff between the disturbance of the quantum states and the acquired 
ability to make inferences about their identity.  We shall also explore the
extent to which the very existence of this tradeoff can be taken as a 
fundamental principle of quantum theory---one from which unitary time evolution 
itself can be derived.

The model we shall base our considerations on is most easily described in terms 
borrowed from quantum cryptography, from whence it takes its origin.  Alice 
randomly prepares a quantum system to be in
either a state $\hat\rho_0$ or a state $\hat\rho_1$.  These states will most
generally be described by $N\!\times\!N$ density operators on an
$N$-dimensional Hilbert space, $N$ arbitrary; there is no restriction that they be pure states, orthogonal, or commuting for that matter.  After the
preparation, the quantum system is passed into a ``black box'' where it may be 
probed by an eavesdropper Eve in any way allowed by the laws of quantum 
mechanics.  That is to say, Eve may first allow the system to interact with an 
auxiliary system, or ancilla, and then perform quantum mechanical measurements
on the ancilla itself \cite{Kraus83}. The outcome of such a measurement may 
provide Eve with some information about the quantum state and may even provide 
her a basis on which to make an inference as to the state's identity.  Upon
this manhandling by Eve, the original quantum system is passed out of the 
``black box'' and into the possession of a third person Bob. 

A crucial aspect of this model is that even if Bob knows the state actually
prepared by Alice and, furthermore, the manner in which Eve operates, because
the system will have become entangled with Eve's ancilla, he will still have to
resort to a new description of the quantum system after it emerges from the 
``black box''---say by some $\hat\rho_0^\prime$ or $\hat\rho_1^\prime$.  This
is where the detail of our work begins.  Eve now has the potential to gather
information about the quantum state; meanwhile that state is no longer a valid
description of Bob's system because it has become entangled with Eve's ancilla.

The ingredients required to formally pose an {\it information/disturbance 
tradeoff principle\/} follow from the details of the model.  We shall need:
\begin{description}
\item[A)]
a concise account of all ancillas and interactions that Eve may use to obtain
evidence about the identity of the state
\item[B)]
a convenient description of the most general kind of quantum measurement she
may then perform on her ancilla
\item[C)]
a measure of the information or inference power provided by any given 
measurement,
\item[D)]
a good notion by which to measure the distinguishability of mixed quantum
states and a measure of disturbance based on it, and finally
\item[E)]
a ``figure of merit'' by which to compare the disturbance with the inference.
\end{description}

In the next section, we shall carry this program out in some detail for two pure
quantum states to give a flavor of how it should go in the general case. 
However, before pursuing this, perhaps we should say a word or two about the
significance of these ideas.  It is often said that it is the Heisenberg
uncertainty relations that dictate that quantum measurements necessarily
disturb the measured system.  To the extent that this statement can be given a 
precise meaning, it is rather beside the point in this context.  The Heisenberg
relations concern the inability to get hold of two classical observables
simultaneously, such as spin in the {\it x\/} direction {\it and\/} spin in
the {\it y\/} direction.  Thus it concerns our inability to ascribe
{\it classical\/} states of motion to quantum mechanical systems.  This, in and
of itself, generally has very little to do with the limits of what can happen to
a quantum state when information is gathered about its identity.  The theme of
this approach differs from that used in formulating the Heisenberg relations in
that it makes no reference to conjugate or complementary variables.  The only 
elements entering these considerations are related to the quantum states 
themselves.  In this way one can get at a notion of state disturbance that is 
purely quantum mechanical, making no reference to the sort of semi-classical
considerations that drove the early conceptual work on the theory's 
interpretation.

What does it really mean to say that the states are disturbed in and of
themselves without reference to classical variables?  It means quite literally 
that Alice faces a loss of predictability about the outcomes of Bob's 
measurements whenever an information gathering eavesdropper intervenes.  Take
as an example the case where $\hat\rho_0$ and $\hat\rho_1$ are nonorthogonal 
pure states.  Then, for each of these, there exists one nontrivial observable
for which Alice can predict the outcome with complete certainty---namely the 
projectors parallel to $\hat\rho_0$ and $\hat\rho_1$, respectively.  However, 
after Alice's quantum states pass into the ``black box'' occupied by Eve, 
neither Alice nor Bob will any longer be able to predict with complete
certainty the outcomes of both these measurements.  This is the real content of 
these ideas.

\section{Pure States}

Let us now carry out the last example in greater detail.  Consider the case
where Alice prepares, with equal probability, one of the following
nonorthogonal states on a two-dimensional Hilbert space ${\cal H}_A$:
\begin{eqnarray}
|0\rangle &=& \cos\alpha\,|a_0\rangle\,+\,\sin\alpha\,|a_1\rangle\\
|1\rangle &=& \sin\alpha\,|a_0\rangle\,+\,\cos\alpha\,|a_1\rangle\;,
\end{eqnarray}
where $|a_0\rangle$ and $|a_1\rangle$ form an orthonormal basis on the space.
A good measure of the nonorthogonality of these states is given by their inner
product:
\begin{equation}
S=\langle 0|1\rangle=\sin 2\alpha\;.
\end{equation}
Eve, in an attempt to gather information about the identity of the state
Alice prepared, will interact Alice's system with an ancilla described by the
states in a Hilbert space ${\cal H}_E$.  This interaction takes its formal
description as a unitary operator $\hat U$ on ${\cal H}_A\otimes{\cal H}_E$.
Supposing the ancilla starts off in some standard pure state $|\psi\rangle$, we
can describe the outcome of this interaction as follows:
\begin{equation}
|s\rangle|\psi\rangle\;\longrightarrow\;
|\Psi^{\scriptscriptstyle {\rm AE}}_s\rangle=
\sum_{n=0}^1\sqrt{\lambda^s_n}\,|A^s_n\rangle|E^s_n\rangle\;,
\end{equation}
$s=0,1$, and where $|\Psi^{\scriptscriptstyle {\rm AE}}_s\rangle$ is written in 
Schmidt polar form \cite{Peres93b} for orthonormal bases (parameterized by
Alice's state $s$) $|A^s_n\rangle$ and $|E^s_n\rangle$ on ${\cal H}_A$ and
${\cal H}_E$ respectively.  The bases and the constants $\lambda^s_n$ will of
course depend on the particular unitary operation $\hat U$ used.  The state of
the system finally left in Eve's possession is given by tracing ${\cal H}_A$
out of the picture, i.e., the mixed state
\begin{eqnarray}
\hat\rho^{\scriptscriptstyle {\rm E}}_s
&=&
{\rm tr}_{\scriptscriptstyle {\rm A}}
\Big(|\Psi^{\scriptscriptstyle {\rm AE}}_s\rangle\langle
\Psi^{\scriptscriptstyle {\rm AE}}_s|\Big)
\nonumber\\
&=&
\lambda^s_0|E^s_0\rangle\langle E^s_0|\,+\,\lambda^s_1|E^s_1\rangle
\langle E^s_1|\;.
\label{RunkleBean}
\end{eqnarray}
Similarly the states passed on to Bob are
\begin{eqnarray}
\hat\rho^{\scriptscriptstyle {\rm A}}_s
&=&
{\rm tr}_{\scriptscriptstyle {\rm E}}
\Big(|\Psi^{\scriptscriptstyle {\rm AE}}_s\rangle\langle
\Psi^{\scriptscriptstyle {\rm AE}}_s|\Big)
\nonumber\\
&=&
\lambda^s_0|A^s_0\rangle\langle A^s_0|\,+\,\lambda^s_1|A^s_1\rangle
\langle A^s_1|\;.
\end{eqnarray}
Eq.~(\ref{RunkleBean}) tells us immediately that we never need consider 
ancillas for Eve with Hilbert space dimension greater than four \cite{Fuchs96b}.
For, at most all four of the $|E^s_n\rangle$ can be linearly independent 
vectors.

With this much as set up, how can we best gauge the tradeoff between 
the disturbance Eve has caused in Alice's system and the information or
inference power she has gained?  There are several ways to go about this
\cite{Fuchs96b}.  For specificity here, we assume that disturbance is gauged
by the average probability that Bob will detect a discrepancy if Alice announces
the state she sent and he checks on the identity of that state.  That is to
say, we will define the disturbance to be
\begin{equation}
D(\hat U)\,=\,1\,-\,\frac{1}{2}\langle 0|
\hat\rho^{\scriptscriptstyle {\rm A}}_0|0\rangle
\,-\,\frac{1}{2}\langle 1|\hat\rho^{\scriptscriptstyle {\rm A}}_1|1\rangle\;.
\end{equation}
(The operator $\hat U$ is listed in this expression to remind us that $D$ is
explicitly a function of whatever unitary operation Eve uses to entangle her
ancilla with Alice's system.)

For a measure of inference power gained by Eve, we take the best possible
average probability that she will make an error in guessing of the state's
identity after performing a measurement on the ancilla.\footnote{Another
measure of Eve's performance is given by the best possible Shannon information
that can be gained about the identity of
$\hat\rho^{\scriptscriptstyle {\rm A}}_s$ \cite{Fuchs96b}.  The present choice
of the error probability in modeling Eve's gains has been made because of its relative tractability---the problem can be handled start to finish without 
numerical simulations---and its consequent illustrative value.}  In order to
get at this expression, one needs the formalism of Positive Operator Valued
Measures \cite{Fuchs96a,Helstrom76}.  However, we record it here for
reference:\footnote{This convenient form of expression was worked out in
collaboration with Jeroen van de Graaf.}
\begin{equation}
P_e(\hat U)\,=\,\frac{1}{2}\,-\,\frac{1}{4}\,{\rm tr}\Bigl|\,
\hat\rho^{\scriptscriptstyle {\rm E}}_1-
\hat\rho^{\scriptscriptstyle {\rm E}}_0\,\Bigr|\;,
\end{equation}
where $|\cdot|$ signifies an operator diagonal in the same basis as its 
argument, but with eigenvalues that are the absolute values of those of the
argument.  It is the tradeoff between $D(\hat U)$ and $P_e(\hat U)$ that we
should like to explore as a function of $\hat U$.

Rather than considering the full-blown problem of evaluating the tradeoff
between these quantities for four-dimensional ${\cal H}_E$, here we stave off
some of the mathematical difficulties encountered there by restricting ourselves
to three-dimensional ${\cal H}_E$.  This allows for an easier path to the
essential physics.  (Besides, in the problem of Ref.~\cite{Fuchs96b} it was
found by numerical simulation that two-dimensional ${\cal H}_E$ always sufficed
for the optimal tradeoff.)  

Let us consider the set of unitary operators that take $|s\rangle|\psi\rangle$
to real superpositions of the basis states $|a_m\rangle|e_\beta\rangle$,
$m=0,1$ and $\beta=x,y,z$, on ${\cal H}_A\otimes{\cal H}_E$.  Also, let us make
the reasonable assumption of only considering $\hat U$s that obtain the same
symmetry on the ancilla's relative states as that of Alice's states in the
sense that:  if
\begin{equation}
\hat U |a_m\rangle|\psi\rangle\,=\,\sum_{n=0}^1|a_n\rangle|R_{mn}\rangle\;,
\end{equation}
then all inner products $\langle R_{m'n'}|R_{mn}\rangle$ remain invariant under
an interchange of $0\leftrightarrow1$ in the indices.\footnote{A more complete
explanation of the origin of these assumptions can be found in 
Ref.~\cite{Fuchs96b}.}

It will be shown in the full paper to be submitted to these proceedings that
the set of all such $\hat U$s can be parameterized by three real numbers
$\lambda$, $\phi$, and $\theta$.  Moreover, $D(\hat U)$ and $P_e(\hat U)$
work out explicitly to be
\begin{eqnarray}
D(\hat U)\!\!
&=&
\!\!\cos^2\!\lambda\left(\sin^2\!\theta\,-\,\frac{1}{2}S\cos2\phi\sin2\theta
\right.
\phantom{\frac{1}{2}S^2(1-\sin}
\nonumber\\
&&
\phantom{\!\!\cos^2\lambda\left(\sin^2\!\theta\right.}
\,+\,\left.\frac{1}{2}S^2\Big(1-\sin2\phi\Big)\cos\!2\theta\right)
\end{eqnarray}
and
\begin{equation}
P_e(\hat U)\,=\,\frac{1}{2}\,-\,\frac{1}{2}\sqrt{1-S^2\,}\sqrt{G(\hat U)}\;,
\end{equation}
where
\begin{equation}
G(\hat U)\,=\,\cos^4\!\lambda\cos^2\!2\phi\,+\,\frac{1}{2}
\sin^2\!2\lambda\Big(1-\sin2\phi\Big)\cos^2\!\theta\;.
\end{equation}

Ideally, what one would like to do with these quantities is form from them a
curve of the smallest possible $D$ as a function of $P_e$.  However that leads
to a pretty hefty constrained-variation problem.  Suffice it for the present to
consider a much simpler problem where the ``figure of merit'' is the ratio
$G/D$---the larger this quantity, the more effective the given $\hat U$ will
have been at leading to a good inference while minimizing the disturbance.
(More details of the full constrained variation problem will appear in the full 
paper.)

For fixed $\theta$ and $\phi$ with $\sin2\phi\ge0$, $G/D$ is maximized
when $\lambda=0$---that is to say, a two-dimensional ancilla again suffices.
Then it is straightforward to calculate the best possible $D$ for a fixed $P_e$.  It is given by
\begin{equation}
D\,=\,\frac{1}{2}\,-\,\frac{1}{2}\!\left\{S^2G\,+\,\left[1-S^2
\left(1-\sqrt{1-G}\,
\right)\right]^{\!2}\right\}^{\!1/2}\;,
\label{Bumfuzzle}
\end{equation}
where
\begin{equation}
G\,=\,\frac{1}{1-S^2}\Big(1-2P_e\Big)^2\;.
\label{GlueBoy}
\end{equation}
Notice that $D\rightarrow0$ as $P_e\rightarrow1/2$, just as one would expect.
$D$ reaches its maximum value when $P_e$ reaches its minimal value.
Eqs.~(\ref{Bumfuzzle}) and (\ref{GlueBoy}) completely define the best possible
inference/disturbance tradeoff in this context.

\section{Mixed States}

The deepest understanding of what the information/dis\-turbance principle has
to say about quantum theory as a whole will necessarily come from the
mixed-state analog of these considerations.  For only then can a direct
comparison be made to classical probability distributions on phase
space---the object of study in classical information theory \cite{Caves96}.
Unfortunately, in moving to the mixed state version of this principle, the
mathematical difficulties become even more acute.  Some progress in the
face of these troubles will be reported in the full paper.

However, in the mean time, there is one restricted result about information
gain vs.\ disturbance for mixed states that brings out an interesting mystery.
This result is known as the ``no-broadcasting theorem'' \cite{Barnum96a}.

Suppose a quantum system, secretly prepared in one state from the set
${\cal A}\!=\!\{\hat\rho_0,\hat\rho_1\!\}$, is dropped into a ``black box''
whose purpose is to {\it broadcast\/} or replicate that quantum state 
onto two separate quantum systems.  That is to say, a state identical
to the original should appear in each system {\it when\/} it is considered
without regard to the other (though there may be correlation or quantum
entanglement between the systems).  Can such a black box be built?  If so, then
that will certainly provide a way to gain information about the mixed state 
without causing a detectable disturbance in the system.

The ``no-cloning theorem'' \cite{Wootters82,Dieks82} insures that the answer
to this question is {\it no\/} when the states in ${\cal A}$ are pure and 
nonorthogonal; for the only way to have each of the broadcast systems described 
separately by a pure state $|\psi\rangle$ is for their joint state to be
$|\psi\rangle\otimes|\psi\rangle$.  When the states are mixed, however, things
are not so clear.  There are many ways each broadcast system can be described
by $\hat\rho$ without the joint state being $\hat\rho\otimes\hat\rho$, the
mixed state analog of cloning. The systems may also be entangled or correlated
in such a way as to give the correct marginal density operators.  

For instance, consider the limiting case in which $\hat\rho_0$ and $\hat\rho_1$
commute and so may be thought of as probability distributions 
$p_0(b)$ and $p_1(b)$ for the eigenvectors in a common diagonalizing basis.  In
this case, one easily sees that the states can be broadcast; the broadcasting 
device need merely perform a measurement of the eigenbasis and prepare two 
systems, each in the state corresponding to the outcome it finds.  The
resulting joint state is {\it not\/} of the form $\hat\rho\otimes\hat\rho$ but
still reproduces the correct marginal probability distributions and thus, in 
this case, the correct marginal density operators.

It turns out that the case just described is indeed a rather special one, for
two states $\hat\rho_0$ and $\hat\rho_1$ can be broadcast
{\it if and only\/} if they commute \cite{Barnum96a}.  That is, if and only
if they can be simultaneously thought of as classical probability distributions
for some underlying reality.  This is the content of the no-broadcasting
theorem.

The enticing mystery that arises from the no-broad\-casting theorem is the
following.  A detailed study of Eqs.~(\ref{Bumfuzzle}) and (\ref{GlueBoy}) and
their analogs in Ref.~\cite{Fuchs96b} reveals that information can be gained
about the identity of two {\it pure\/} quantum states without disturbance if
and only if they are orthogonal.  (Also see Ref.~\cite{Bennett92a}.)  It
follows that there can be information gain without disturbance if and only if 
the two pure states can be cloned.

One might expect the same sort of behavior to hold for mixed states:  that
information can be gained without disturbance if and only if the two states
can be broadcast.  This thought, however, is misguided.\footnote{I thank Ben
Schumacher and Michael Nielsen for discussion on these points.}  Life becomes 
interestingly more complex when it comes to mixed states.

A simple example to
consider is that of two density operators on a four-dimensional Hilbert space.  
In matrix representation, suppose they do {\it not\/} commute on a 
$2\!\times\!2$ block but do on the orthogonal complement to it.  The states will thus have two common eigenvectors.  Suppose
the eigenvalues associated with the common eigenvectors are all distinct.
Then it follows immediately that there are information gathering measurements 
that will cause no disturbance to these states---it consists of checking whether the system is in the $2\!\times\!2$ block or either of the other two 
eigenvectors.  This measurement can be performed as a quantum nondemolition
measurement.  Since the eigenvalues are all distinct, there are distinct
probabilities for all these outcomes and thus information to be gained.
However, because these two states are noncommuting overall, they cannot be 
broadcast.

An interesting open question is the necessary and sufficient mathematical
criteria required of two density operators to insure that information gathering
measurements necessarily disturb the quantum states.  A conjecture is that this
will be the case when they commute on no nontrivial subspace.

\section{Foundations}

Finally we briefly sketch the statement of a theorem that indicates that the
information/disturbance tradeoff principle may be a fundamental aspect of
quantum theory.\footnote{I thank Michel Boyer, Gilles Brassard, Nicolas Gisin,
Asher Peres, and Bill Wootters for listening to me patiently on these points.}
Details of its proof will be given in the full conference paper.

Let us go back to the Hilbert space ${\cal H}_A\otimes{\cal H}_E$ and suppose
we consider all of quantum mechanics intact and beyond question except for the
general group of time evolutions that this space may be submitted.  We shall
suppose that this group ${\cal T}$ consists of all maps (continuous in time)
that are bijections of ${\cal H}_A\otimes{\cal H}_E$ onto itself, and that it
contains {\it at least\/} all the unitary operations.  A priori, however, we
may not wish to tie down the set of evolutions any further than this---it might 
contain other linear maps, nonlinear maps, maps discontinuous with respect to
the topology of the vector space, or what have you.

As it stands, not much can be said about the group ${\cal T}$.  So let us now
consider what would be required of a mapping $\Phi\in{\cal T}$ if it were to be
capable of breaking the principles espoused in this paper.  A mapping $\Phi$
is said to allow {\it illegal eavesdropping\/} (i.e., information gain without
disturbance) if there are two
nonorthogonal states $|s\rangle$, $s=0,1$, in ${\cal H}_A$ and a standard
state $|\sigma\rangle$ in ${\cal H}_E$ such that
\begin{equation}
\Phi\Big(|s\rangle|\sigma\rangle\Big)=|s\rangle|\sigma_s\rangle\;,
\end{equation}
$s=0,1$, for which $0\le|\langle\sigma_0|\sigma_1\rangle|<1$.  (All vectors in
this description are assumed normalized.)  If such a map existed, then---since
we are assuming all the other principles of quantum mechanics are still
intact---a measurement on ${\cal H}_E$ alone will certainly reveal information
about the state $|s\rangle$ without disturbing it.

Suppose now, we find the existence of such maps too unbearable, and we take
it as a principle that such time evolutions cannot exist.  The question is, can
we still be left with time evolutions that are more general than those provided
by the unitary group?  For instance, one can easily imagine maps on
${\cal H}_A\otimes{\cal H}_E$ that are perfectly well behaved on product states,
doing just what we expect, even though their behavior goes completely awry
on the set of entangled states.

The answer to the question is ``no.''  If there is a map in ${\cal T}$ that
allows the increase or decrease of the modulus of {\it any\/} inner product of
two states in ${\cal H}_A\otimes{\cal H}_E$, then we can construct another map
that will use that effect for illegal eavesdropping.  This will be shown in the
full paper.  Thus, if the information/disturbance tradeoff principle is to
be upheld, ${\cal T}$ can only contain inner-product modulus preserving maps.
And this, it turns out, is the premise for Wigner's famous theorem 
\cite{Peres93b} stating that, allowing the possible redefinition of phase, all 
such maps must be unitary or anti-unitary.  If the maps are to be continuous in 
time, then they must be unitary.

This simple point demonstrates that there are things to be learned about quantum
theory itself by observing how it can be used for communication and computation.
The result is not completely satisfactory in that we had to require that
${\cal T}$ contain at least the unitary maps before we could make any progress.
Nevertheless, it does provide food for thought.

\end{document}